# Tidal debris morphology and the orbits of satellite galaxies

David Hendel,[1]* Kathryn V. Johnston[1]
[1] *Department of Astronomy, Columbia University, 550 W 120th St., New York, NY 10027, USA*

**ABSTRACT**

How do galaxies move relative to one another? While we can examine the motion of dark matter subhalos around their hosts in simulations of structure formation, determining the orbits of satellites around their parent galaxies from observations is impossible except for a small number of nearby cases. In this work we outline a novel approach to probing the orbital distributions of infalling satellite galaxies using the morphology of tidal debris structures. It has long been understood that the destruction of satellites on near-radial orbits tends to lead to the formation of shells of debris, while those on less eccentric orbits produce tidal streams. We combine an understanding of the scaling relations governing the orbital properties of debris with a simple model of how these orbits phase-mix over time to produce a 'morphology metric' that more rigorously quantifies the conditions required for shells to be apparent in debris structures as a function of the satellite's mass and orbit and the interaction time. Using this metric we demonstrate how differences in orbit distributions can alter the relative frequency of shells and stream structures observed around galaxies. These experiments suggest that more detailed modeling and careful comparisons with current and future surveys of low surface brightness features around nearby galaxies should be capable of actually constraining orbital distributions and provide new insights into our understanding of structure formation.

**Key words:** galaxies: kinematics and dynamics – dark matter – galaxies: haloes – galaxies: statistics

## 1 INTRODUCTION

In the modern cosmological picture large galaxies are built up over time by the hierarchical merging of many smaller projenitor systems (e.g. White & Rees 1978). Since the galactic luminosity function has a (truncated) power-law slope (Schechter 1976, and many others) we expect that the vast majority of these mergers will have one parent that is significantly more massive than the other. Cosmological simulations have shown that Milky Way-mass galaxies have typically experienced less than one major merger (with mass ratio $\xi \geq 0.1$) since redshift $z = 1$ (Fakhouri, Ma & Boylan-Kolchin 2010) and so minor mergers are an important part of their total mass accretion rate at late times (Oser et al. 2010), although possibly subdominant to 'diffuse' accretion of intergalactic gas (Fakhouri & Ma 2010).

Observationally, the appearance of debris from minor mergers can be broadly divided into two morphological categories. Stream-like substructures stretch approximately along the progenitor's orbit, sometimes wrapping around the host multiple times, for example as seen in the recent deep imaging of NGC 5907 by Martínez-Delgado et al. (2008). Shell-like structures, like those in the vast complex around NGC 474 (Duc et al. 2013), may extend significantly both along and perpendicular to the path of their disrupted parent, forming an umbrella-shaped distribution of stars and/or sharp edges in the light distribution, frequently interleaved with each other.

Interest in the detailed study of debris structures has been stimulated over the last two decades in part by the discovery of numerous streams around the Milky Way, including the tidal tails of of Palomar 5, a globular cluster (Odenkirchen et al. 2001), the stream of the Sagittarius dwarf galaxy (Majewski et al. 2003), the GD-1 stream (Grillmair & Dionatos 2006), and the Orphan stream (Grillmair 2006; Belokurov et al. 2007). The formation of tidal streams is well understood, which makes them powerful tracers of the host galaxy's potential and history: individual stars, heated either by tidal forces from the host or internal two-body interactions in the case of globular clusters, leave the satellite through the Lagrange points that mark saddle points in the the system's effective potential. Stars that leave through the inner Lagrange point have lost energy relative to the satellite and have slightly shorter orbital periods, causing them to stretch out over time into a leading tail, while particles that have gained energy have longer orbital periods and lag behind, forming a trailing tail. This model can give estimates of the width and length of streams given the orbit, progenitor mass and interaction time, and analytic representations of this picture have been used to successfully produce realistic stream models without full N-body simulations (Johnston, Sackett & Bullock 2001; Küpper, Lane & Heggie 2012; Sanders 2014; Gibbons, Belokurov & Evans 2014; Bovy 2014; Amorisco 2015; Fardal, Huang & Weinberg 2014).

* E-mail: hendel@astro.columbia.edu

While there is yet to be entirely convincing evidence of shell structures around the Galaxy despite several candidates (such as the Triangulum-Andromeda and Hercules-Aquila clouds, Rocha-Pinto et al. 2004; Belokurov et al. 2006; Deason et al. 2013), they are an important class of features in external galaxies. Perhaps the original description of substructure as a shell occurs in the Atlas of Peculiar Galaxies (Arp 1966), in reference to Arp 230. The Palomar 200-inch plates show at least six shells perpendicular to the principal axis. Many other galaxies in the Atlas have shell-like debris structures, e.g. Arp 223 and 227, an early hint that they are common - although some processing may be necessary to reveal them due to their low surface brightness (Malin & Carter 1983). Simulations of shells followed, starting with restricted n-body models where the host potential was static and the disrupting satellite's self-gravity was assumed negligible (Quinn 1984; Dupraz & Combes 1986). Subsequent work modeled the satellite potential self-consistently (Piran & Villumsen 1987; Heisler & White 1990). Both showed that radial mergers can reproduce the observed shell properties. Observational results disfavor models where the shells are formed through the production of density waves in the stellar outskirts of the host galaxy by a perturber (the Weak Interaction Model, Thomson & Wright 1990) as an alternative to the merger scenario in many systems (e.g. Turnbull, Bridges & Carter 1999; Wilkinson et al. 2000; Schiminovich, van Gorkom & van der Hulst 2013). More recent theoretical treatments of shell formation have focused on recovering the host gravitational potential from the line-of-sight velocity distribution of the debris (Merrifield & Kuijken 1998; Ebrová et al. 2012; Sanderson & Helmi 2013). Detailed models can capture the shell density distribution for purely radial orbits (Sanderson & Bertschinger 2010) but no simple description of the conditions under which minor merger debris forms a shell rather than a stream has emerged until very recently (Amorisco 2015, this work).

Overall, previous studies demonstrate that merger debris' properties are functions of the interaction time (length of streams or angular extent and number of shells) and mass ratio (which influences the spread of the debris, both in position space and energetically). Since the disruption process is well enough understood to actually be invertible (e.g. Johnston 1998; Helmi & White 1999), the population of streams with different extents and surface brightnesses can conceptually be used to ascertain the rate of minor mergers of different mass ratios (as outlined in Johnston, Sackett & Bullock 2001; Johnston et al. 2008) and place more stringent constraints on structure formation than a coarser classification as major or minor mergers. However, the signatures of most accretion events are expected to be extremely low surface brightness (LSB), starting near 28 mag arcsec$^{-2}$, with the majority below $\sim$ 30 mag arcsec$^{-2}$ (Johnston et al. 2008; Cooper et al. 2010); fortunately, exploration of the LSB universe is rapidly advancing with the availability of wide-field cameras, specialized instruments, and LSB-optimized observing techniques (Janowiecki et al. 2010; Miskolczi, Bomans & Dettmar 2011; Atkinson, Abraham & Ferguson 2013; Martínez-Delgado et al. 2010; Abraham & van Dokkum 2014; Duc et al. 2015). The future Large Synoptic Survey Telescope (LSST) is expected to image the Southern sky with surface brightness sensitivity 5 magnitudes deeper than current wide-field surveys such as the Sloan Digital Sky Survey (SDSS) (Ivezic et al. 2008).

The promise of these current and near-future investigations motivates a reexamination of the utility of substructure as probes of galaxies' pasts. Since the debris structures that result from accretion events with near-radial infall are morphologically distinct from those produced along less eccentric orbits, substructure counting could conceptually be sensitive to the accretion rates as a function of mass, time *and* orbit - adding a new dimension of information about the accretion history that is otherwise difficult to access. It is not possible to reconstruct the full orbits of most objects from observations since only the projected distance and line-of-sight component of motion can be measured outside the Local Group and the typically small number of known galaxy-mass satellites in individual systems (outside clusters) precludes the use of statistical tools such as the Jeans equations without stacking (Herbert-Fort et al. 2008). While each galaxy may have only one or fewer debris structures apparent, a well-constructed survey of galaxies to faint surface brightness could provide constraints on the collective distribution of infalling objects. A strong grasp on the orbital distribution of accreting objects provides an important next step towards reducing the number of degrees of freedom in the choice of cosmological models, similar to the way that the merger fraction as measured by galaxy pairs has been used to restrict combinations of the principle cosmological parameters (Carlberg 1991; Conselice et al. 2014), and might have implications for our understanding of how baryons inhabit low-mass haloes.

In this paper we explore the idea that debris structures carry useful information about the satellite orbital infall distribution. In Section 2, we briefly describe N-body simulations used to examine debris formation throughout the rest of this work. In Section 3 we extend the simple stream-building picture described above to also encompass shells and use the results to define the morphology metric $\mu$. In Section 4 we use our metric to demonstrate that the population of shells and streams is indeed sensitive to the infalling subhalos' orbital velocity distribution and that these populations could potentially be used to provide interesting constraints on cosmological models that predict different orbital infall distributions, presuming a variety of intermediate challenges can be overcome. We consider the observational and modeling advances necessary to reasonably apply this technique in Section 5. Section 6 concludes.

## 2 N-BODY SIMULATIONS

We performed a set of dark-matter-only (but see 4.2, where baryonic effects are considered) N-body simulations with the self-consistent field basis function expansion code (Hernquist

& Ostriker 1992) to explore the formation of debris structures across a wide range of orbital and galactic parameters. In each simulation, a $10^5$ particle NFW-profile (Navarro, Frenk & White 1996) satellite was inserted at the apogalacticon of its orbit in a static, NFW host potential. The satellite evolves first in isolation to ensure that it equilibrates and then the host potential is gradually turned on over 10 satellite internal dynamical times to reduce artificial gravitational shocking. Total energy is conserved to ∼1% of the satellite internal potential energy during the 8 Gyr integrations. The simulation is not halted if the satellite is completely disrupted since the debris continues to evolve.

The satellites are initialized with masses $m/\mathrm{M}_\odot = 6.5 \times 10^6, 6.5 \times 10^7, 6.5 \times 10^8$, and $6.5 \times 10^9$, where $m$ is the mass enclosed at 35 NFW scale radii, out to which particles were realized. The scale radius $r_0$ was adjusted for each mass so that the mean density inside the scale radius is the same for each, with a value of $r_0 = 0.86$ kpc for the $6.5 \times 10^9$ $\mathrm{M}_\odot$ satellite. This scaling gives them the same fractional mass loss rate along a given orbit.

We use a host that is broadly consistent with expectations for a Milky Way-scale dark matter halo, choosing a viral mass of $1.77 \times 10^{12}$ $\mathrm{M}_\odot$, virial radius of 389 kpc and a scale radius of 24.6 kpc. We chose a set of orbits with total energy equal to that of circular orbits at 25, 50, and 100 kpc in this potential and varied the angular momentum in twelve steps between 0.05 and 1.0 times the angular momentum of the circular orbit for each energy, denoted by $L_{\mathrm{circ}}$. The lowest angular momentum case for these orbits is quite radial; the $r_{\mathrm{circ}} = 25$ kpc, $L/L_{\mathrm{circ}} = 0.05$ simulation has a perigalacticon distance of $\sim 0.65$ kpc and an apogalacitcon of $\sim 40$ kpc. This orbit is a reasonable match to the extent of the shell systems seen around NGC 4651 (Foster et al. 2014) and MGC-5-7-1 (Schiminovich, van Gorkom & van der Hulst 2013), both of which have dark matter halo masses estimated to be of the same order as our selected value.

These simulations were used to check the scaling relations in Section 3.2 and to substantiate a choice of the morphology metric that will separate shells from streams.

## 3 RESULTS I: DEFINING A MORPHOLOGY METRIC

To build a statistic that will allow us to quickly assess what debris from a disrupted satellite will look like given a particular set of merger parameters we first presume that the host potential is sufficiently well known (at the population-statistics level) that the fact that a shell forms at all places a constraint on the characteristics of an individual merger. This is in contrast to previous work (Merrifield & Kuijken 1998; Ebrová et al. 2012; Sanderson & Helmi 2013) that uses shells to constrain host halo parameters. Those analyses concentrated on modeling shell systems around individual galaxies in detail through density and line-of-sight velocity matching, while here we are interested in the general conditions that lead to shells and streams (without assuming any spectral information) in order to examine the implications of observed merger morphologies for the distribution of satellite orbital properties around a large sample of galaxies.

In this section we begin by reviewing the properties of test particle orbits in extended mass distributions, then use these insights to understand what properties of mergers lead to the creation of shells and streams (Section 3.1). Identifying the debris' energy and angular momentum dispersions $\sigma_{\mathrm{E}}$ and $\sigma_{\mathrm{L}}$ as the quantities of interest in the morphological distinction, we describe a method to estimate the dispersions in Section 3.2. Finally we fold this understanding into a morphology metric that can predict the debris morphology in Section 3.3.

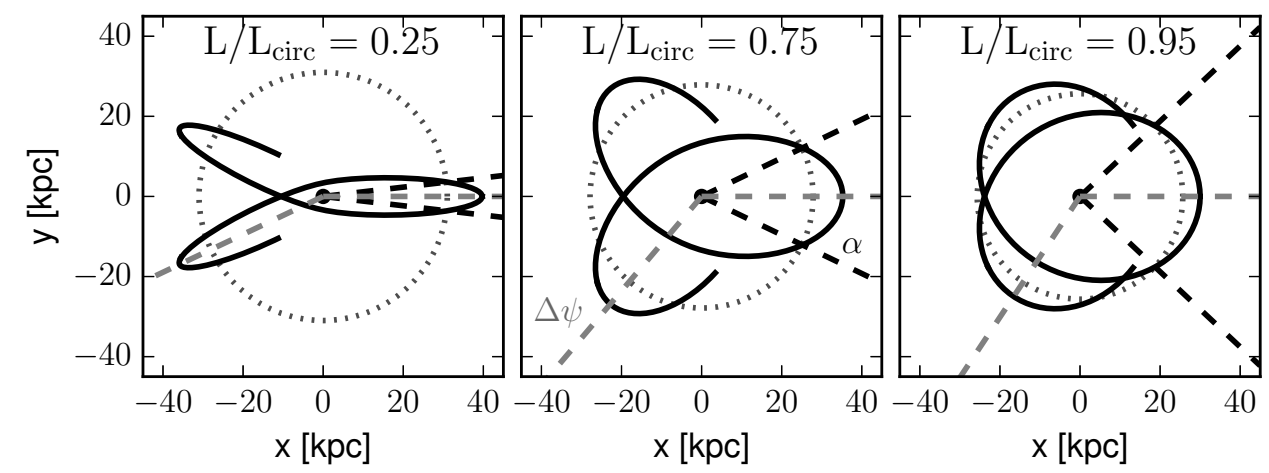

**Figure 1.** A visualization of the angles $\alpha$ (black dashed lines) and $\Delta\psi$ (gray dashed lines) for orbits of varying eccentricity. The solid black lines in each panel show an orbit with energy equal to the 25 kpc circular orbit but with the specified fraction of that orbit's angular momentum. The dotted circle marks the half-period radius; a particle will spend equal time inside and outside this radius. For the orbits shown it varies from 25.7 kpc for the most circular orbit to 31 kpc for the most eccentric orbit. Inside $\alpha$ a particle is closer in time to apogalacticon than perigalacticon.

### 3.1 Properties of orbits

Knowledge of the orbits of test particles in simple potentials is an important first step towards understanding mergers in their entirety, as shown in the original classic work by Toomre & Toomre (1972). In the case of debris resulting from a dwarf galaxy entering a Milky Way-size galaxy's halo we can begin by assuming that the remnant's gravitational influence on the unbound particles is negligible. This assumption is well justified for mass ratios smaller than $\xi = m/M_{host} = 10^{-4}$ (Choi, Weinberg & Katz 2007) and in our simulations the debris' conserved quantities are unaffected by the satellite after unbinding until $\xi \gtrsim 10^{-3}$. Additionally, we expect the host potential at the typically large radii considered below to be dominated by the host's (spherical) monopole term, since any effect from disks must be subdominant given that both streams and shells are found in galaxies with elliptical morphologies as well. While deviations from spherical symmetry will introduce a variety of interesting and potentially important effects such as orbital plane precession and chaos, as a limiting case spherical hosts are useful for building intuition.

Orbits in spherical potentials can be uniquely identified by their conserved quantities - the total energy E and total angular momentum L - up to the orbital plane orientation, which we take as the $x - y$ plane for convenience. These, combined with the potential parameters, determine the shape and properties of the orbit: the radii of the turning points at apogalacticon and perigalacticon, radial orbital period $T_r$, and the precession per orbit between apogalactica $\Delta\psi$. Fig. 1 illustrates how orbits change when the circularity $L/L_{\mathrm{circ}}$ is varied. The precession angle and perigalac-tic distance decrease with decreasing circularity while the apogalactic distance increases. We will also find it useful to consider the angular 'width' of a single petal of the rosette traced by an orbit, which varies systematically with angular momentum; near-radial orbits spend the vast majority of their time near apogalacticon, covering only a small angle in azimuth during a large fraction of their radial period, while near-circular orbits may cover hundreds of degrees in azimuth during the same portion of radial phase. To characterize this difference, we define $\alpha$ as the angle through which a particle moves during the outer half of its radial period. This is a proxy for the angle subtended by the orbit's minor axis, which does not exist for these non-closed orbits.

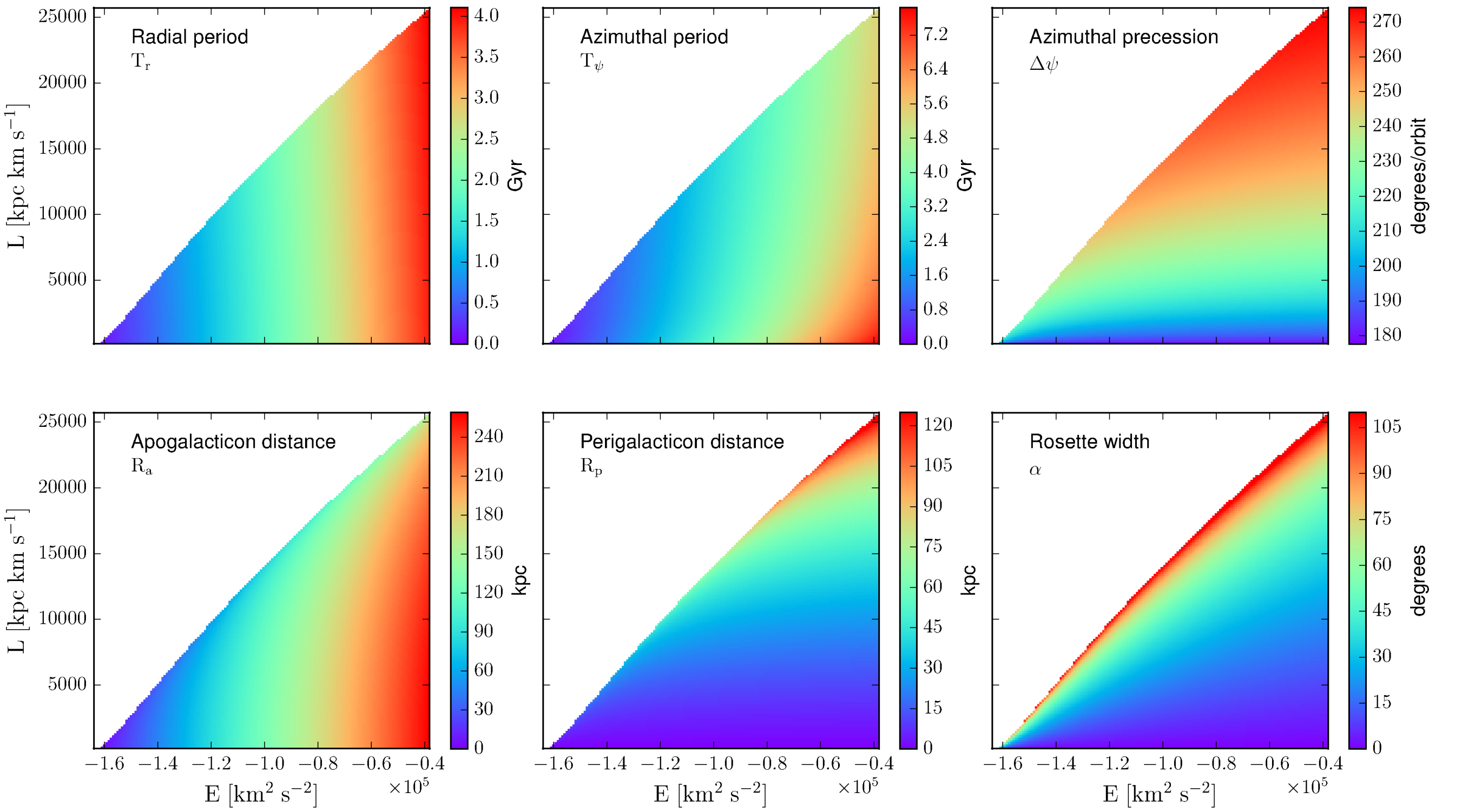

**Figure 2.** Various orbital descriptors as a function of energy and angular momentum for test particles orbiting in the spherical NFW host halo used for the N-body simulations (Section 2). White space indicates regions that are inaccessible in this host potential. The periods and apogalacticon distance are strong functions of energy but weak functions of angular momentum, whereas the changes in precession and perigalacticon distance are dominated by angular momentum. The "width" $\alpha$ is primarily determined by $L/L_{circ}$.

Fig. 2 illustrates how the orbital properties vary as a function of energy and angular momentum. Each panel shows the available space for bound orbits with energies less than that of a circular orbit at 150 kpc in the same host as described in Section 2 - an NFW halo with virial mass $1.77 \times 10^{12}$ $M_\odot$ and a scale radius of 24.6 kpc. The radial period $T_r$ is a strong function of energy but only weakly dependent on angular momentum, while for the precession per orbit $\Delta\psi$ the opposite is true. Apogalacticon and perigalacticon distances are also primarily functions of energy and angular momentum, respectively, but the dependencies are weaker here than for $T_r$ and $\Delta\psi$. The angle $\alpha$ follows neither of these trends but instead depends on the orbit's circularity rather than the absolute values of $E$ or $L$.

Fig. 3 shows the visible effects of sampling the properties shown in Fig. 2 across a small range in the $(E, L)$ parameter space. The gray rosette represents the position of a reference orbit at every time step of a simple numerical leapfrog integration in the standard halo described above; it has energy equal to that of a circular orbit at 37 kpc but only 30% as much angular momentum, i.e. $L/L_{circ} = 0.3$. The points near the rosette show the positions of a few orbits with a distribution of properties around the reference orbit at a few different times. Ten particles start at (x,y,z) = (25,0,0) kpc and for the left (right) panel their energies (angular momenta) were varied by ±2.5% (±10%) in equal spacings. The energies and angular momenta set the magnitude and direction of the initial particle velocities and their positions are evolved with the same leapfrog scheme. As expected from Fig. 2, varying the orbital energy while holding L constant primarily affects the orbital periods of particles, spreading them out in radial phase but keeping their orbits otherwise similar for cosmologically significant interaction times (here 4 Gyr), giving them the appearance of a stream tracing the 'progenitor' orbit. The most obvious deviation from the reference orbit is an increased apogalacticon distance for the trailing particles and decreased apogalacticon distance for the leading ones; this also results in the rosette petal tips, as traced by the debris, becoming more rounded than the reference orbit as high-energy particles move farther out when they are near apogalacticon. In contrast, when angular momentum is varied but energy is held constant particles remain tightly clustered in radial phase because their periods are nearly equal. However, they slowly spread azimuthally due to differences in the rate at which their apogalactica precess, resulting in a 'shell' of particles at the same radius near apocenter since the apogalactic distance varies only weakly with L. The selected values of the conserved quantities and halo do not change the qualitative effect.

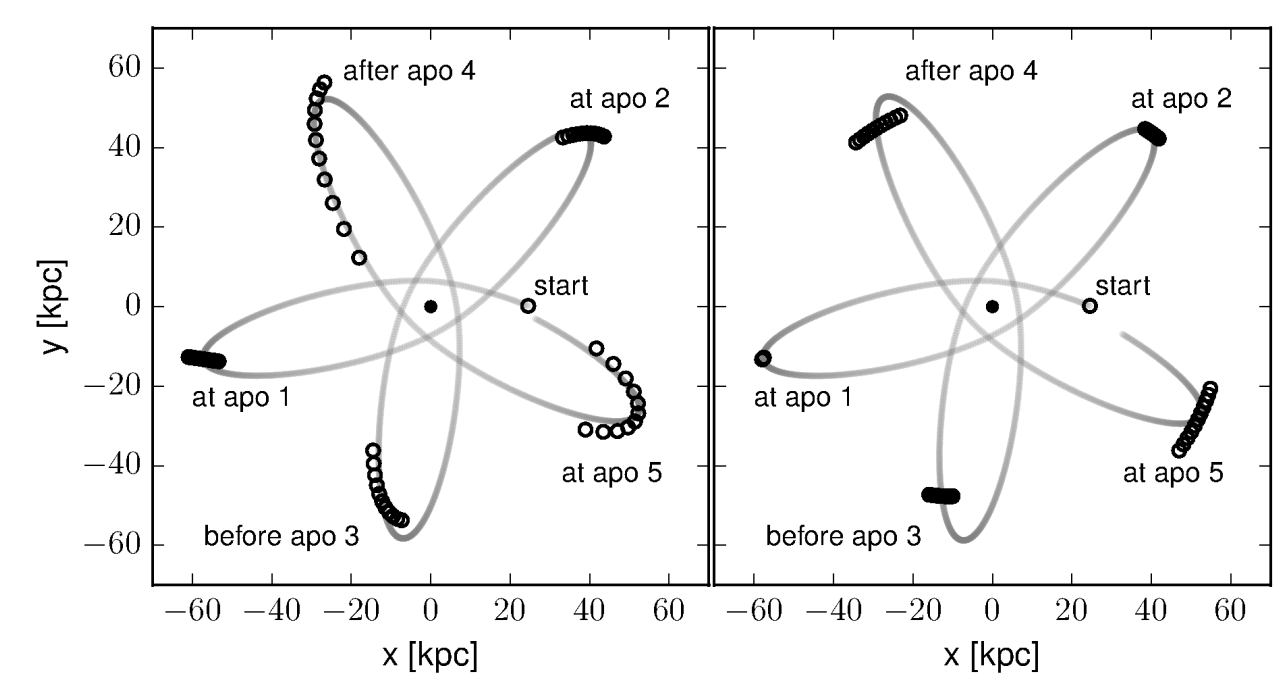

**Figure 3.** Variation of test-particle orbits due to small changes in energy (left) or angular momentum (right); points show the positions of particles that are displaced from from the gray central orbit in each quantity by $\pm 2.5\%$ ($\pm 10\%$). Both figures use the same reference orbit and are integrated in the same potential and points are displayed at the same times. While variation in E displaces particles primarily along the reference orbit, varying L causes precession that distributes particles in azimuth with a range in position angle that grows with time.

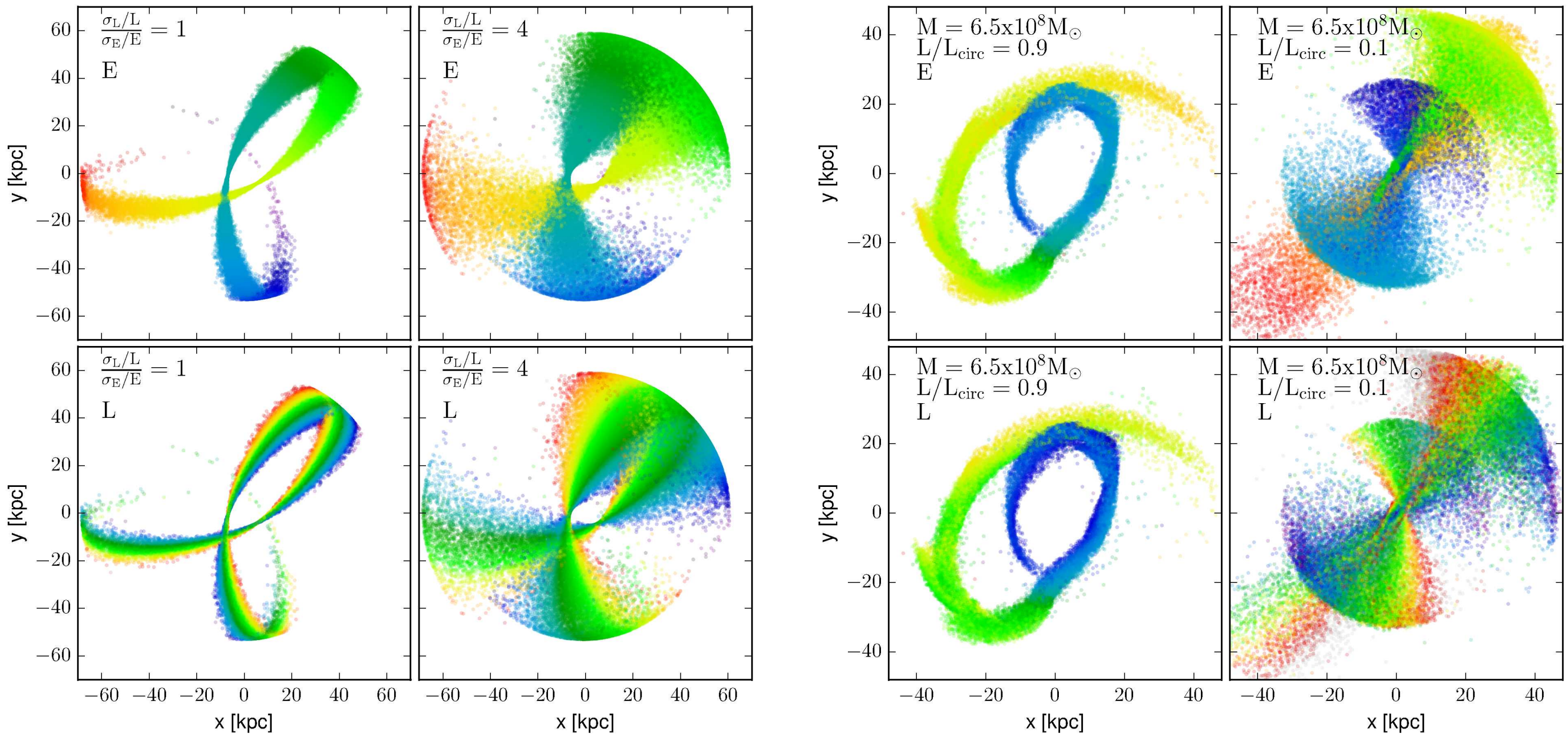

**Figure 4.** Debris formation resulting from simple test particle models for the satellite (left) and and N-body simulations for the same (right). The background potential has a static NFW profile. Individual particles are color coded by either their energy (top row) or angular momentum (bottom row) to illustrate how offsets in orbital quantities dictate their final spatial position.

In reality, streams and shells will contain particles offset from the satellite in both energy and angular momentum. The left panels of Fig. 4 show the result of integrating $10^5$ test particles on the same orbit as Fig. 3 but where the particles are given independent Gaussian distributions in energy and angular momentum, with dispersions $\sigma_E$ and $\sigma_L$ respectively. The energy dispersion is 2% of the mean orbital energy in all panels. The particles are color-coded by their energy (top panels) or angular momentum (bottom panels), with blue indicating the lowest values and red the highest. This simple setup produces quite reasonable 'debris' (compare with the full N-body simulations in the right panels of Fig. 4), which is perhaps surprising, but this model is essentially a minimum-complexity version of the phase-space-distribution method to produce streams (Küpper, Lane & Heggie 2012; Gibbons, Belokurov & Evans 2014; Amorisco 2015). As before the qualitative result of energy and angular momentum sorting is independent of the specific orbit or values of $\sigma_E$ and $\sigma_L$ chosen.

Fig. 4 also shows that when the fractional variation in E and L is equal, $(\sigma_L/L)/(\sigma_E/E) = 1$ (left panels), the distribution has a clearly stream-like appearance. Particles of the same energy (i.e. same color in upper left panel) have very nearly the same period, reaching apogalacticon at the same time. The precession rate, dominated by variation in angular momentum, spreads the particles in azimuth and creates the sharp, rainbow-colored edge at the ends of the rosette petals (lower left panel). The precession is insufficient to significantly alter the appearance that the debris follows the primary orbit. When $(\sigma_L/L)/(\sigma_E/E) = 4$, however, the precession dominates and the debris appears as a shell because the angle subtended by the radial edge due to differences in the precession rate exceeds the width of the rosette petals by a factor of $\sim 3$.

From previous work it is well known that shells are more frequent in the aftermath of radial mergers. Here we have additionally demonstrated that, for a given orbit, the ratio of fractional dispersions in E and L inside a debris structure is an important factor in determining the morphology. However, $\sigma_E$ and $\sigma_L$ are not free parameters but are instead set by the orbit and the internal properties of the satellite.

As shown in Section 3.2 below and Fig. 5, the actual orbital parameter distribution in the debris a) is bimodal, b) has a specific, quite non-Gaussian shape in each mode, and c) has significant covariance especially for low eccentricity orbits (see also their projections into action-angle and frequency space in Sanders & Binney 2013; Bovy 2014; Fardal, Huang & Weinberg 2014). This covariance is why the left panels of the test panel integration in Fig. 4 have a more pronounced apogalacticon edge than real streams: the E-L covariance will only allow a small subset of angular momenta at a given energy, and therefore smaller differences in integrated precession angle at constant E. We find that accounting for this covariance is unnecessary to predict the morphology from N-body simulations (Section 3.3)

### 3.2 Debris scaling relations

The typical scales in debris (which characterize the dispersions $\sigma_{\rm E}$ and $\sigma_{\rm L}$) can be computed following previous work (e.g. Johnston 1998; Binney & Tremaine 2008). Consider a satellite moving with velocity $V_p$ at its perigalacticon, where its distance from the host galaxy is $R_p$. Most mass loss will occur near pericenter so we anticipate that the debris' dispersions in E and L are set by the relevant satellite scales there. Particles are very likely to leave the satellite through the effective Lagrange points where the effective potential has a saddle point; these points are in general not equally spaced but are colinear with the host and satellite centers. We approximate their location relative to the satellite center by the tidal radius $r_{tide}$ where the gravitational forces would balance if the galaxies were point masses on circular orbits:

$$r_{\rm tide} = \left(\frac{m}{3M(R_p)}\right)^{1/3} R_p = sR_p \tag{1}$$

where $m$ and $M(R_p)$ are the mass of the satellite and the mass of the host enclosed inside $R_p$, respectively.

We take energy scale $e_s$ of the debris to be the difference in the host's gravitational potential energy across the satellite diameter at pericenter using the simple linear approximation

$$e_s = 2r_{\rm tide} \left.\frac{\partial \Phi}{\partial R}\right|_{R_p} \tag{2}$$

while the angular momentum scale is computed from

$$l_s = \Delta L = \Delta V R + V \Delta R. \tag{3}$$

Using the satellite internal velocity dispersion $\sigma = \sqrt{Gm/r_{\rm tide}}$ for $\Delta V$, $\Delta R = 2r_{\rm tide}$, and $V_p = \sqrt{GM/R_p}$, one finds

$$l_s = \sigma R_p + 2V_p r_{\rm tide} = (\sqrt{3}+2)sL \tag{4}$$

Thus $e_s$ and $l_s$ are specified by the orbital and galactic parameters and therefore, given our conclusions in Section 3.1, so is the debris morphology; varying $e_s$ and $l_s$ independently for a given satellite and orbit as in the test particle integration of Fig. 4 is unphysical.

The efficacy of these scales is demonstrated in Fig. 5. Shown are the results of the simulations described in Section 2 for $r_{\rm circ} = 25$ kpc with the m $= 6.5 \times 10^6 {\rm M}_\odot$ run in red and m $= 6.5 \times 10^7 {\rm M}_\odot$ in blue, after 8 radial periods (approximately 4.3 Gyr). In each panel satellite particles are

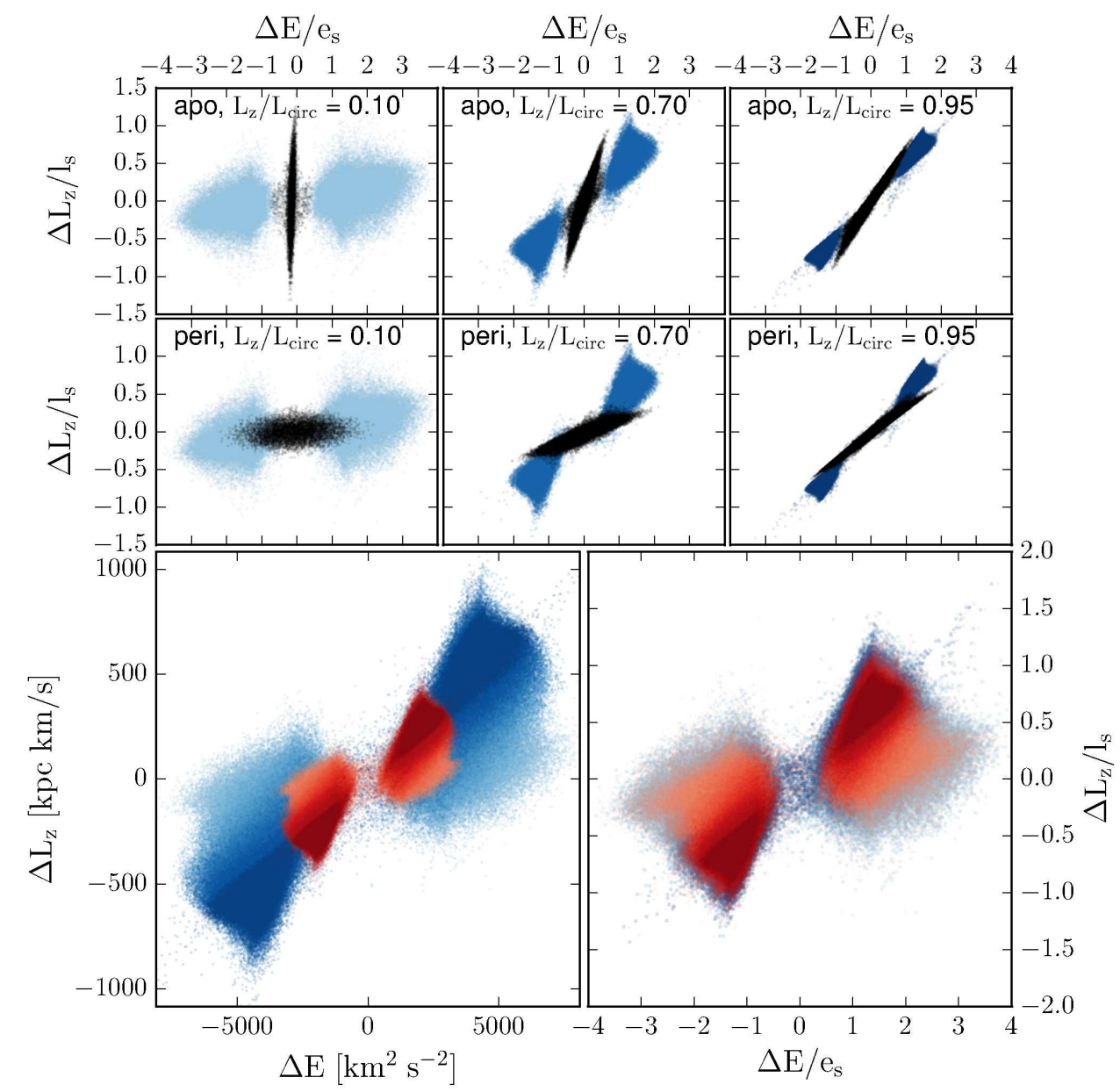

**Figure 5.** Energy and angular momentum offsets of bound and unbound particles.

plotted in $(E, L)$ space as offsets from the satellite center of mass values $E_{orb}$ and $L_{orb}$; all but the bottom-left panel have their axes rescaled by $e_s$ and $l_s$. In the upper panels black points indicate particles that are still bound to the satellite while colored points have become unbound through tidal stripping. The color tone (lightness) of colored points represents the angular momentum of the satellite orbit they come from, with darker tones indicating more circular orbits. Snapshots of the disruption near apocenter (top row) and pericenter (middle row) for a range of angular momenta are provided; the bound particle distribution rotates about the satellite center ($E_{orb}$,$L_{orb}$) between these directions, filling the space with unbound particles (outside an exclusion zone defined by the satellite's interaction with its own debris). The energy and angular momentum of individual particles is constant to high precision after they have been unbound for a short time, consistent with the assumption of no self-interaction at these mass ratios ($\xi \sim 10^{-5} - 10^{-6}$)

In the bottom row, the unbound particles from simulations with a wide range of satellite angular momenta $\rm L/L_{circ} = 0.1, 0.2$ ... 0.9 are plotted together for the two different mass satellites. Shown are absolute offsets (left) and rescaled offsets (right). Although there is some remaining angular momentum dependence the scalings do an excellent job of removing the mass dependence and reducing the variation in median offsets across circularities. While corrections could be made to $e_s$ and $l_s$ to account for the remaining angular momentum dependence and other effects such as mass loss over time, dynamical friction, the changing in tidal radius as a function of instantaneous galactocentric distance, or more general potentials, we consider these simple estimates sufficient for our purpose.

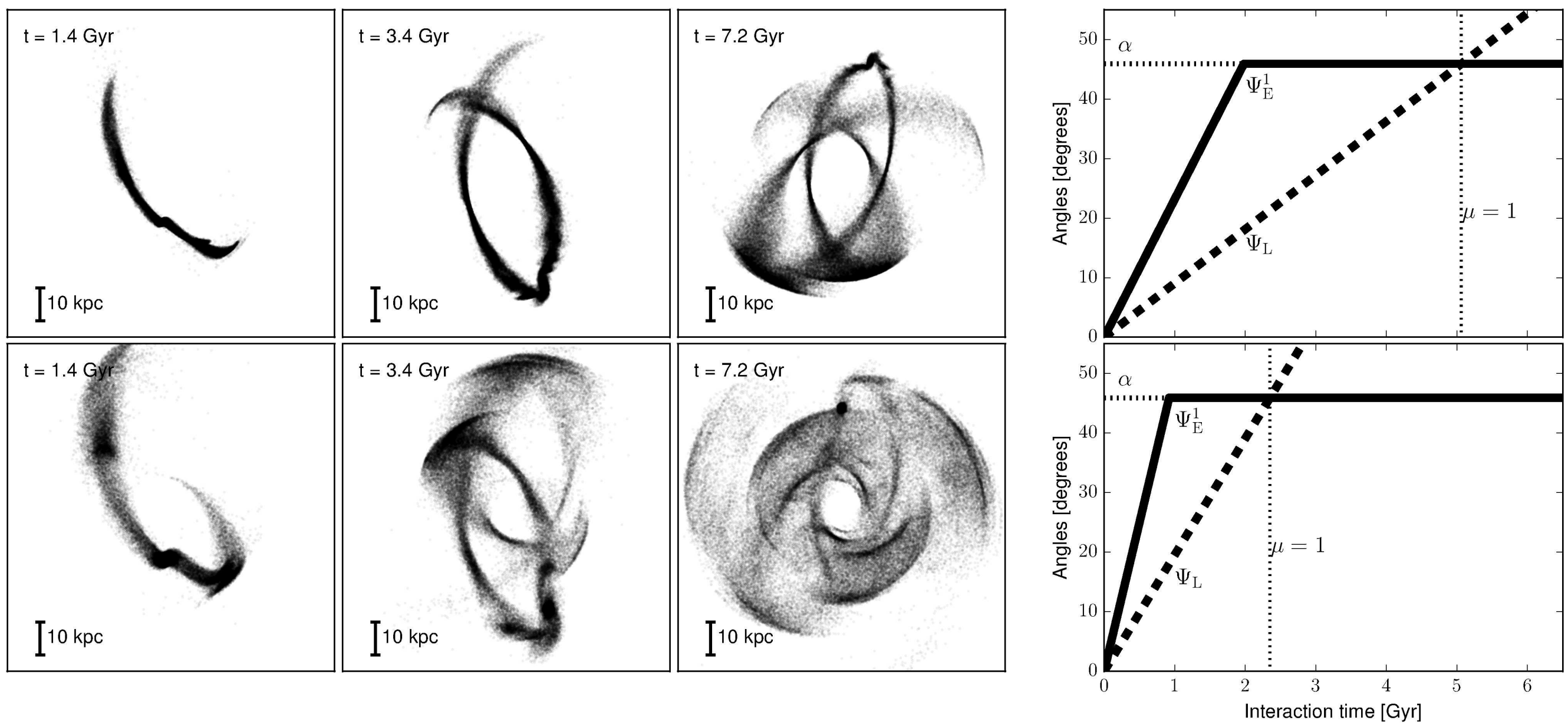

**Figure 6.** Time evolution of debris. Left: the simulation panels show snapshots from the $r_{circ} = 25$ kpc, $L/L_{circ} = 0.7$ simulations of satellites with $m = 2.5 \times 10^7 M_\odot$ (top row) and $m = 2.5 \times 10^8 M_\odot$ (bottom row) at selected times, viewed along the axis perpendicular to the orbital plane.. Right: the various angles, corresponding to the displayed simulations, that contribute to the morphology metric.

### 3.3 A morphology metric

So far we have established that (i) modest variations in orbital parameter space have effects on properties (orbital periods, precession) that are important to the shape of debris on the timescale of satellite disruptions, (ii) which properties will be affected by such variations depend strongly on whether it is E or L that is varied, and (iii) satellite disruption produces debris in a predictable section of orbital parameter space in the vicinity of the satellite's own orbit.

Using these facts, we propose a metric for estimating whether a debris structure would be visually classified as a shell or a stream by comparing the contribution of energy-dominated and angular momentum-dominated effects to its shape. As illustrated in Fig. 4, a reasonable choice for the factor determining morphological assignment is whether or not the shell edge subtends a larger range of position angle than a single rosette petal. We therefore need a measure of 1) the rosette petal angle, already defined as $\alpha$, 2) the angle through which debris has spread due to differences in azimuthal precession rate as a function of time, $\Psi_L$ and 3) the angle subtended due to stream-like orbital period variations in the debris, $\Psi_E$, if less than a whole petal.

The near-complete separability of effects from $E$ and $L$ allows a simple linear approximation. Given a satellite-host pair and an orbit $(E_{orb}, L_{orb})$, the angular size of the constant-energy edge caused by precession of apogalactica is just the differential precession per orbit with respect to $L$ at $E_{orb}$ times the angular momentum scale and the number of orbits $N_{\rm orb}$:

$$\Psi_L \equiv l_s N_{\rm orb} \left. \frac{\partial \Delta\psi}{\partial L} \right|_{E_{orb}}, \quad (5)$$

where the contribution proportional to $\partial\Delta\psi/\partial E$ is neglected due to the weak dependence of $\Delta\psi$ on $E$. Similarly, since the radial period is a weak function of $L$ the debris' extent from energy variations can be approximated by considering the added (subtracted) angle that short- (long-) period, low- (high-) energy debris moves through, given by

$$\Psi_E \equiv e_s N_{\rm orb} \left. \frac{\Delta\psi}{T_r} \frac{\partial T_r}{\partial E} \right|_{L_{\rm orb}}. \quad (6)$$

We find that $\Psi_E > \Psi_L$ generically; dynamically young debris will always be stream-like. We restrict $\Psi_E$ to be less than $\alpha$ so that we are considering individual rosette petals (see Fig. 1) and so we define our morphology metric $\mu$ as

$$\mu \equiv \frac{\Psi_L}{\Psi_E^1} \quad (7)$$

where

$$\Psi_E^1 \equiv \min(\alpha, \Psi_E). \quad (8)$$

Fig. 6 illustrates the time evolution of the angles $\Psi_E$ and $\Psi_L$ as computed from Equations (5) and (6) for two mergers in our standard host halo with $r_{circ} = 25$ kpc and circularity $L/L_{circ} = 0.7$ but differing masses. When $\mu$ is larger than 1 shells should become apparent in the debris. The rosette petal width $\alpha$ (horizontal dotted line) is a function of orbit only and therefore constant with respect to $\xi$ and time. The stream angle $\Psi_E^1$ (solid black line) is larger than $\Psi_L$ (dashed black line) at early times; both angles increase linearly so $\mu(t)$ is constant until $\Psi_E$ exceeds $\alpha$. Eventually $\Psi_E > \alpha$ and so $\Psi_E^1 = \alpha$ thereafter. Since the integrated precession angle $\Psi_L$ continues to grow, the morphology becomes increasingly dominated by angular momentum effects until $\Psi_L = \Psi_E^1$, i.e. $\mu = 1$ (vertical black dotted line), when the merger has produced a shell. At the last snapshot both mergers show shells as expected from to the moderately eccentric orbit but the mass dependence results in a significantly faster transition for the higher mass case, by more than a factor of two in this example. The time dependence of $\mu$ adequately captures the morphological evolution.

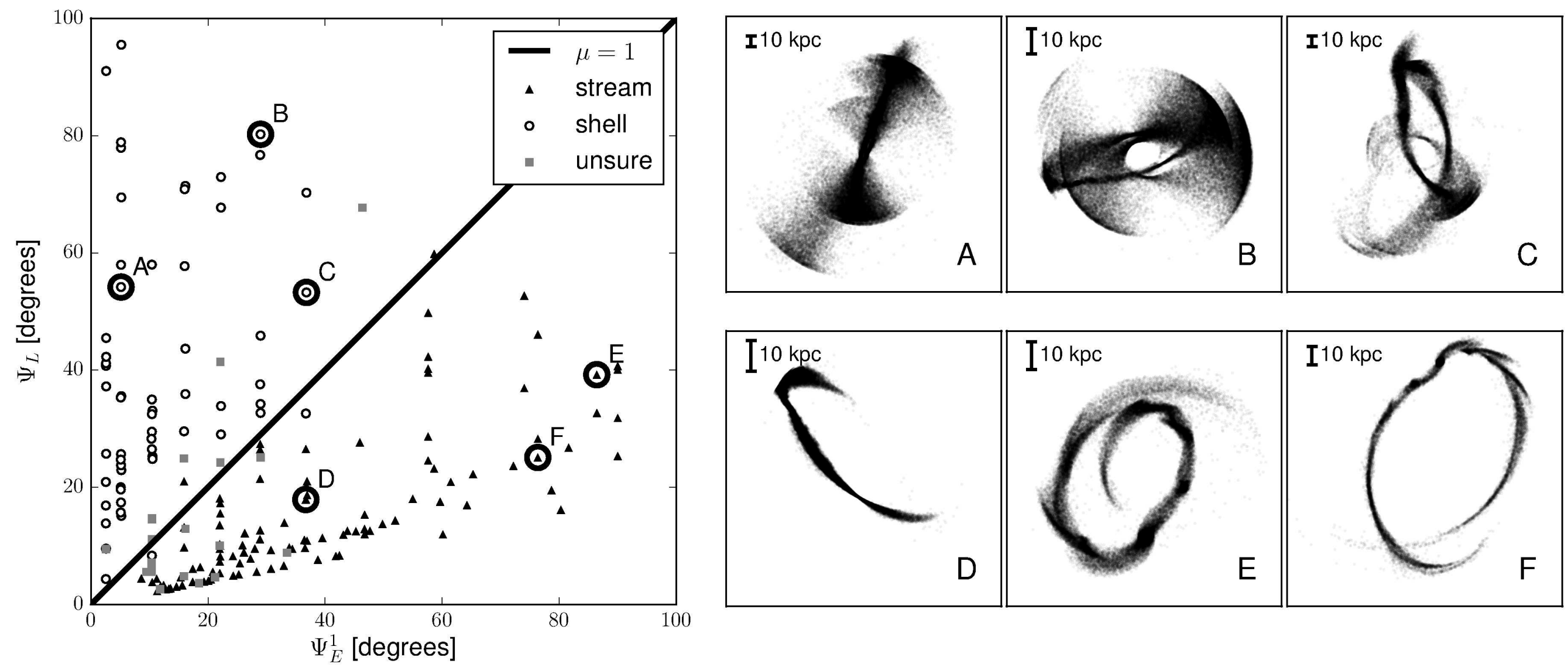

**Figure 7.** Left: the theoretically motivated divider $\mu = 1$ (black line) effectively separates debris structures that have been visually classified as containing shells (red circles) from those marked as streams (blue triangles). Most objects that could not be confidently assigned were dynamically young. Some objects with $\Psi_L > 100°$ have been omitted for clarity. Right: selected objects with $\mu = 10.6$, 2.8, 1.4, 0.49, 0.45, and 0.33, viewed along the axis perpendicular to the orbital plane.

In this model all debris structures inevitably evolve into shells. At late times phase mixing will reduce the density enhancements to the point where debris may not be identified as such and it might be desirable to exclude mergers that have progressed to this stage from our accounting in order to fairly compare with observations; however, this process is closely tied to related issues of surface brightness and orientation effects that are left for future work.

To further test the morphology metric, 200 snapshots were selected at random from the 14,400 outputs of the simulations described in Section 2 and were morphologically classified by eye, viewed along the axis perpendicular to the orbital plane. Equations (5) and (8) were then used to compute the energy- and angular momentum-induced position angle variations from the simulation initial conditions. Fig. 7 shows that debris marked as containing shells versus those that appear as only streams are cleanly separated in ($\Psi_L$, $\Psi_E^1$) space, and that $\mu = 1$ can effectively divide the different structures. In the right-hand panels, six representative snapshots are shown with a wide range of $\Psi_L$, $\Psi_E^1$, and $\mu$, illustrating how sorting by decreasing values of the morphology metric shifts from classical shell systems (A, B) to "umbrellas" (C) and then streams (D, E, F).

## 4 RESULTS II: THE INFLUENCE OF ORBITAL DISTRIBUTIONS ON THE FREQUENCY OF SHELLS

The rapid acceleration of computational speed, storage, and techniques now allows cosmological N-body simulations of structure formation to evolve representative volumes of the Universe to $z = 0$ in order to track statistical samples of haloes down to dwarf galaxy scales, $\mathrm{M}_{200} \sim 10^{10}\mathrm{M}_\odot$ (for example, Millennium-II, Boylan-Kolchin et al. 2009; Bolshoi, Klypin, Trujillo-Gomez & Primack 2011). Of particular interest for our study is the haloes' accretion history. Typically the accretion history is projected into two spaces: the *mean merger rate* $\mathrm{dN_m/d\xi/dz(\xi, M, z)}$ (which has units of mergers per halo, for a halo mass M, per unit redshift z per unit mass ratio $\xi$) and the *orbital infall distribution* which is usually described by the probability distribution $\mathrm{P(V_r, V_\theta \mid \xi, M, z)}$ for finding satellites with radial and tangential velocities in the range $\mathrm{V_r + dV_r}$ and $\mathrm{V_\theta + dV_\theta}$.

The morphology metric $\mu = \mu(M, \xi, z, E, L, \Phi(M, z))$ described in Section 3.3 allows us to make a direct connection between the orbital infall distribution and the observed numbers of shells and streams. The expected number of mergers that produce debris structures exhibiting shells around a galaxy of mass M can be computed, after relating the orbital parameters to their equivalent velocities through the assumed halo potential $(E, L) \mapsto \Phi(V_r, V_\theta)$, by integrating over the distributions

$$N_{\mathrm{shell}}(M) = \int_{z_{\mathrm{min}}}^{z_{\mathrm{max}}} \int_{\xi_{\mathrm{min}}}^{\xi_{\mathrm{max}}} \int_{V_\theta=0}^{\sqrt{V_{\mathrm{esc}}^2 - V_r^2}} \int_{V_r=0}^{V_{\mathrm{esc}}} \frac{\mathrm{d}N_m}{\mathrm{d}\xi\mathrm{d}z} \times \\ \mathrm{P}(V_r, V_\theta \mid \xi, M, z)\mathrm{H}(\mu - \mu_t)\mathrm{d}V_r\mathrm{d}V_\theta\mathrm{d}\xi\mathrm{d}z. \quad (9)$$

Here $V_{\mathrm{esc}}$ is the host escape velocity, $\mu_t$ is the value of the morphology metric chosen to demarcate the transition from streams to shells, and $\mathrm{H}(x)$ is the Heaviside step function, equal to 1 if $x > 0$ and 0 otherwise. Computing $N_{\mathrm{stream}}(M)$ instead simply requires negating the argument of H. The quantities $N_{\mathrm{shell}}(M)$ and $N_{\mathrm{stream}}(M)$ are the model predictions for the absolute number and relative frequency of mergers that create each type of debris structure. It is important to note that $N_{\mathrm{shell}}$ counts the number of events that produce shells, not the number of individual surface brightness edges resulting from a merger, and therefore there is risk of observational confusion when multiple progenitors produce debris around the same host galaxy. Additionally, it is assumed that all events inside the integration limits are detectable without regard to uncertainties introduced by the host light distribution and any survey sensitivity effects. Nevertheless, this analytic approach allows us to make some first estimates of how sensitive the population of shells and streams might be to the orbital infall distribution. Section 4.1 outlines the current state of measurements of various properties of haloes in cosmological simulations. In Section 4.2 we implement an approximate representation for the luminous matter in the satellites and effective surface brightness limits. Combining the two, Section 4.3 makes predictions for the observed debris population.

### 4.1 Parameter and model choices

We presume that at some future date a merger rate, NFW halo concentration relation, and orbital infall distribution will all be consistently measured from a single cosmological simulation. To schematically illustrate how these predictions could be compared to observations we will use some commonly cited fits from separate simulations, each of which necessarily has its own cosmological parameters, cosmic variance, systematic errors and resolution effects to consider. Any other choice for each may be trivially substituted.

#### *4.1.1 Merger rate*

The per-halo merger rate provides the overall scaling of the absolute number of substructures. Fakhouri, Ma & Boylan-Kolchin (2010) combined the wide Millennium-I (Springel et al. 2005) simulation with the better mass resolution of Millennium-II (Boylan-Kolchin et al. 2009) to achieve good statistics on mergers over 5 decades of host mass and merger mass ratio from the present to z $\approx$ 15. They provide a fitting function to the mean merger rate per halo $\mathrm{d}N_m/\mathrm{d}\xi/\mathrm{d}z$ as a separable function of host mass M, mass ratio $\xi$ and redshift z:

$$\frac{\mathrm{d}N_m}{\mathrm{d}\xi \mathrm{d}z}(M,\xi,z) = A\left(\frac{M}{10^{12}\mathrm{M}_\odot}\right)^{\alpha}\xi^{\beta}\exp\left[\left(\frac{\xi}{\bar{\xi}}\right)^{\gamma}\right](1+z)^{\zeta} \tag{10}$$

where $(\alpha,\beta,\gamma,\zeta) = (0.133,-1.995,0.263,0.0993)$, $(A,\bar{\xi}) = (0.0104, 9.72\times10^{-3})$. The rate per halo is a weak function of host mass and redshift but a much stronger function of mass ratio; the quality of satellite and host masses measured in the observations will have a large effect on the uncertainty in $N_{\rm shell}$.

#### *4.1.2 Host halo parameters*

As indicated by Equation 2 and also shown by Amorisco (2015), the slope of the host halo's potential at perigalacticon directly affects the spread in energy and hence the rate of shell formation. For spherical NFW haloes of a given mass this slope is determined by the concentration parameter $\mathrm{c_{vir}} = \mathrm{r_{vir}}/\mathrm{r_s}$, where $\mathrm{r_{vir}}$ is the host virial radius and $\mathrm{r_s}$ is the NFW scale radius. Navarro, Frenk & White (1996) showed that there is a tight correlation between concentration and mass; recently Dutton & Macciò (2014) found the following relation holds for relaxed haloes with $10^{10} < \mathrm{M/M_\odot} < 10^{15}$ to redshift z = 5 with a simulation using the cosmological parameters as measured by the *Planck* satellite (Planck Collaboration et al. 2014):

$$\begin{aligned} &\log_{10}\, c_{\rm vir} = a + b\, \log_{10}(M/10^{12}h^{-1}\mathrm{M}_\odot) \\ &a = 0.537 + (1.025 - 0.537)\exp(-0.718z^{1.08}) \\ &b = -0.097 + 0.024z. \end{aligned} \tag{11}$$

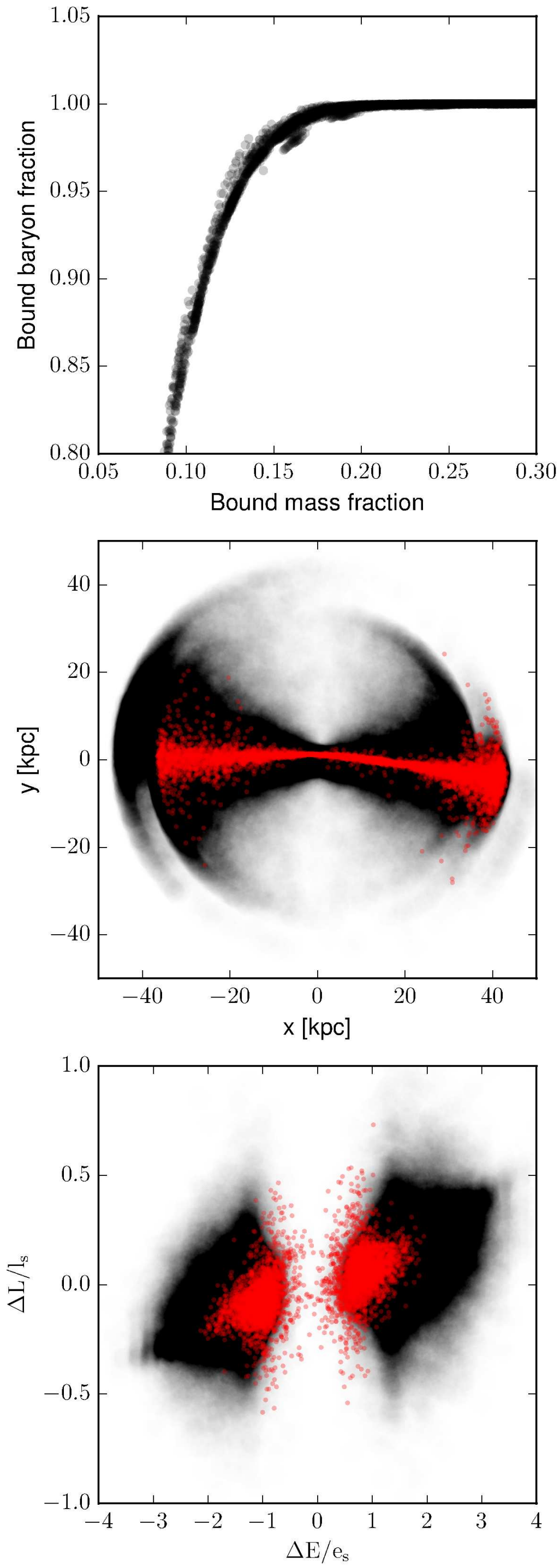

**Figure 8.** Outside-in stripping of mass. Top: "baryons" represented by the particles with initial binding energies in the 90th percentile and higher, are protected from tides until $\sim$ 80% of the total mass has been lost independent of the mass, orbital energy, and orbital angular momentum of the satellite. Center: Shells in dark matter (black) vs baryons (red). Bottom: The baryons (red) have smaller energy and angular momentum offsets because they are only stripped when the satellite has lost a large fraction of its mass (compare the bottom-left panel of Fig. 5)

#### *4.1.3 Orbital parameter distributions*

The distribution of orbital parameters may depend on a number of merger properties: the host mass, the mass ratio, and the redshift of infall are all likely candidates besides the chosen cosmology. Past work has disagreed about which of these are important. Some suggest $\xi$ dependence (Tormen 1997; Vitvitska et al. 2002; Jiang et al. 2015) while others do not (Gill et al. 2004; Khochfar & Burkert 2006; Wetzel 2011). Some find redshift dependence (Benson 2005; Wetzel 2011) while others who investigated exclude it (Zentner et al. 2005). Whether the host mass has any influence is also disputed, however there are large variations in dynamic range between these studies which may obfuscate this. The authors who investigated correlation between orbital parameters typically found it (Tormen 1997; Benson 2005; Wang et al. 2005; Jiang et al. 2015).

We consider possible correlation between $\mathrm{V_r}$ and $\mathrm{V_\theta}$ an important factor since it is required e.g. to capture the situation where infalls have a preferred energy. With this in mind we will use the findings of Benson (2005) and Jiang et al. (2015) to test whether the number of shells and streams can distinguish between infall models and therefore, given sufficient grips on what range of M, $\xi$, and z one is probing, the consistency of simulation and observation. Benson (2005) provides fits to the probability distribution $\mathrm{P(V_r, V_\theta \mid z)}$ for $\mathrm{z} \in \{0.0, 0.5, 1.0\}$ while the results of Jiang et al. (2015) can be transformed to compute $\mathrm{P(V_r, V_\theta \mid M, \xi)}$ with $\mathrm{M/M_\odot} \in \{10^{12}, 10^{13}, 10^{14}\}$ and $\xi \in \{[0.0001, 0.005], [0.005, 0.05], [0.05, 0.5]\}$. The shapes of these infall distributions in $(V_r, V_\theta)$ space are shown in the left panels of Fig. 11 and Fig. 12

### 4.2 Converting dark matter simulations to LSB features

#### *4.2.1 Two-component satellites*

Thus far we have considered the disruption of only the satellites' dark component. The baryonic part is situated deep in the satellite potential and is therefore significantly more resistant to tides than the more extended dark matter. This can lead to substantial differences in the spatial distribution of the two components simply because the satellite will have lost much of its mass before any stars are unbound and therefore the area covered by stars in $(E, L)$ space will smaller than computing $e_s$ and $l_s$ from the initial mass would suggest. We represent the fraction of the binding energy distribution that contains stars by $f_b$. The true value of this fraction is not well understood and depends on the details of the galaxy formation process but previous work has shown that tagging the 10% most bound dark matter particles with stars reproduces many properties of the stellar halo in high-resolution resimulations of Milky Way-sized galaxies (De Lucia & Helmi 2008) and captures the observed surface brightness, velocity dispersion, and luminosity distributions of Local Group dwarfs modeled from semi-analytic initial conditions (Bullock & Johnston 2005) although a smaller fraction (1-3%) may be required to ensure that the size-luminosity relation of Local Group dwarfs is also respected when the halo is live (Cooper et al. 2010).

Fig. 8 explores the most extreme two-component scenario, where all particles with initial binding energy in the 90th percentile and above are considered 'baryons,' $f_b = 0.1$. In the top panel the fraction of baryon-labeled particles that are still bound is plotted as a function of the total bound mass fraction at each output for 72 of our simulations, those with $L/L_{\rm circ} \leq 0.5$; this cut is used to ensure sufficient number statistics. As expected the baryons are well isolated from tidal forces and virtually none of them are removed until $\sim$ 80% of the initial mass is stripped. This result matches that found previously by Villalobos et al. (2012) and Chang, Macciò & Kang (2013). The correlation between the remaining fractions of the two components is very similar across all simulations and the deviations that exist do not appear to be related to mass ratio, circularity or orbital energy.

This multiple component model has two important effects on the morphology metric. First, there is a time delay introduced by the requisite halo stripping; this can be less than an orbital period if the orbit is highly eccentric or more than a Hubble time for near-circular orbits at the virial radius where the tidal field is comparatively weak and the mass loss rate is slow. Since the delay depends sensitively on $f_b$, the mass-to-light ratio and the profile of the embedded stars we simply note that where it is large is also the regime where debris will be streams for tens of Gyr and therefore $N_{\rm stream}$ may be better interpreted as an upper limit since some subhalos will not yet have developed visible tidal features. Second, since the angles $\Psi_E$ and $\Psi_L$ are proportional to $m^{1/3}$ through the energy and angular momentum scales their rate of increase is slowed proportionally. The orbital width $\alpha$ is not affected and the result $\Psi_E > \Psi_L$ still holds, therefore shell development is slowed by a factor of $\sim (0.20)^{-1/3} = 1.7$.

#### *4.2.2 Integration limits: approximating observability*

To facilitate a first test of the sensitivity of morphological fractions to orbit distributions, we make a number of assumptions regarding which mergers will be visible and counted as shells or streams, leaving more detailed estimates to future work.

- The threshold $\mu = 1$ is assumed to divide the morphological classes exactly, independent of orientation.
- We presume that the existence of coherent debris structures implies that there has not been a major merger (which we take to be mergers greater than 10:1) during disruption and therefore and set $\xi_{\rm max} = 0.1$.
- Host mass growth is neglected, consistent with the previous assumption.
- The surface brightness and therefore detectability of substructures are a function of mass, orbit and time (Johnston et al. 2008; Cooper et al. 2010). As an estimate we assume that debris structures from $z_{\rm max} = 0.5$ ($\sim$ 5 Gyr) will still have high enough surface brightness to be visible if the progenitor was sufficiently massive, choosing sufficiently to be $\xi_{\rm min} = 0.05$ for convenient comparison with measured infall distributions. Infalls that occur at higher redshift or with smaller mass ratios are not counted. A simple extension would be a mass-ratio-dependent maximum infall redshift; a full accounting would require more complex analysis to assess the influence of debris morphology and orientation on the lifetime of observability.

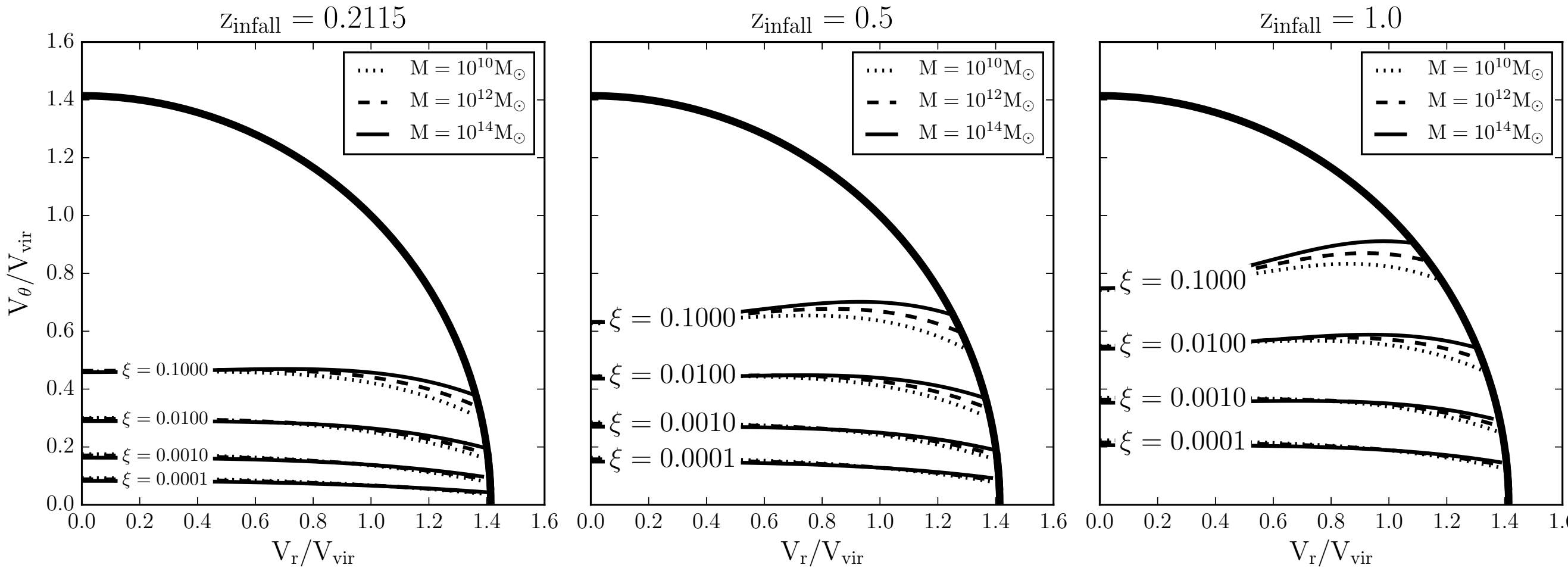

**Figure 9.** Slices through the four-dimensional space that Equation 9 integrates over. The outer black line indicates the host halo's escape velocity while the twelve interior lines show where $\mu = 1$, i.e. the transition from shells to streams, in different merger scenarios. Any satellite whose orbit places it below the relevant line at the given redshift, host mass, and mass ratio will create debris with a shell morphology. The dependence on host mass at fixed $\xi$ is due to the mass-concentration relation changing the slope of the host potential at perigalacticon and therefore affecting $e_s$.

- The minimum infall redshift for which debris may be extended enough to be noticeable is a function of orbital period and depends also in $f_b$ through the time delay discussed in Section 4.2 but for simplicity we choose $z_{\rm min} = 0.1$ ($\sim$ 1.3 Gyr). Another way to define this constraint would be to use $z_{\rm min} = 0$ but only count mergers with $\max(\Psi_E^1, \Psi_L) > \Theta$ where $\Theta$ is some minimum angle; for example, a few times the angle subtended by $r_{\rm tide}$ at apogalacticon.
- Motivated by the results of Section 4.2 we compute the scales $e_s$ and $l_s$ using 20% of the initial satellite mass to capture the effect of stripping the extended dark matter before any stars are removed.
- Finally, we assume that dynamical friction and asphericity negligibly alter the orbital parameters during the erosion of the extended dark halo.

All of these effects are worthy of further investigation.

### 4.3 Sensitivity to orbital infall distributions

In this section we explore how much influence infall distributions have on observed number of galaxies that contain shells and streams. First we will consider two arbitrarily chosen distributions $P(V_r, V_\theta)$ that are widely separated and have simple functional forms (Section 4.3.1). Having established this baseline, we consider the differences that result from two similar infall distributions measured in the same simulation with the same technique but at different redshifts (Section 4.3.2) and finally to mass dependence in the distributions (Section 4.3.3) using the indicator $N_{\rm shell}(M)$.

Evaluating the four-dimensional integral in Equation 9 numerically is difficult because computing the derivatives of orbital quantities that are used to calculate $\mu$ requires four optimizations and seven integrations. Applying an adaptive quadrature method in four dimensions would require tens of millions of evaluations and is therefore impractical, so instead we use a significantly modified and parallelized version of MCINT[1], a Monte Carlo integrator, to reach acceptable $\sim$ 1% accuracy with a few tens of thousands of evaluations.

One way to visualize the division of orbits into shell-forming and stream-forming is to consider the $(V_r, V_\theta)$ plane and draw lines of $\mu = 1$ (the selected value for the morphological transition) for different mass ratios and host masses at a single redshift. This is shown in Fig. 9. The infall distribution is then a probability density function (pdf) of the coordinate axes and the fractional ratio of shells and streams at the given $(M, \xi, z)$ is the integral of that pdf on either side of the $\mu = 1$ line.

#### *4.3.1 A simple test - distinguishing infall distributions*

As a proof of concept, we use Equation 9 to compute the expected fraction of galaxies that would be observed with shell or stream debris under the assumptions of Section 4.2.2 given the two simple, Gaussian infall distributions shown in $(V_r, V_\theta)$ space in the upper panel of Fig. 10. The first, displayed with red dashed contours, is axis-aligned and centered on $(V_r, V_\theta) = (0.7, 0.4)$; this set of orbits is radially biased. The second distribution, shown in solid blue contours, is centered on $(V_r, V_\theta) = (0.75, 0.75)$ and oriented at a 45° angle to the axes which produces orbits that have higher average energy and a smaller dispersion in energy than the first distribution.

[1] https://pypi.python.org/pypi/mcint/

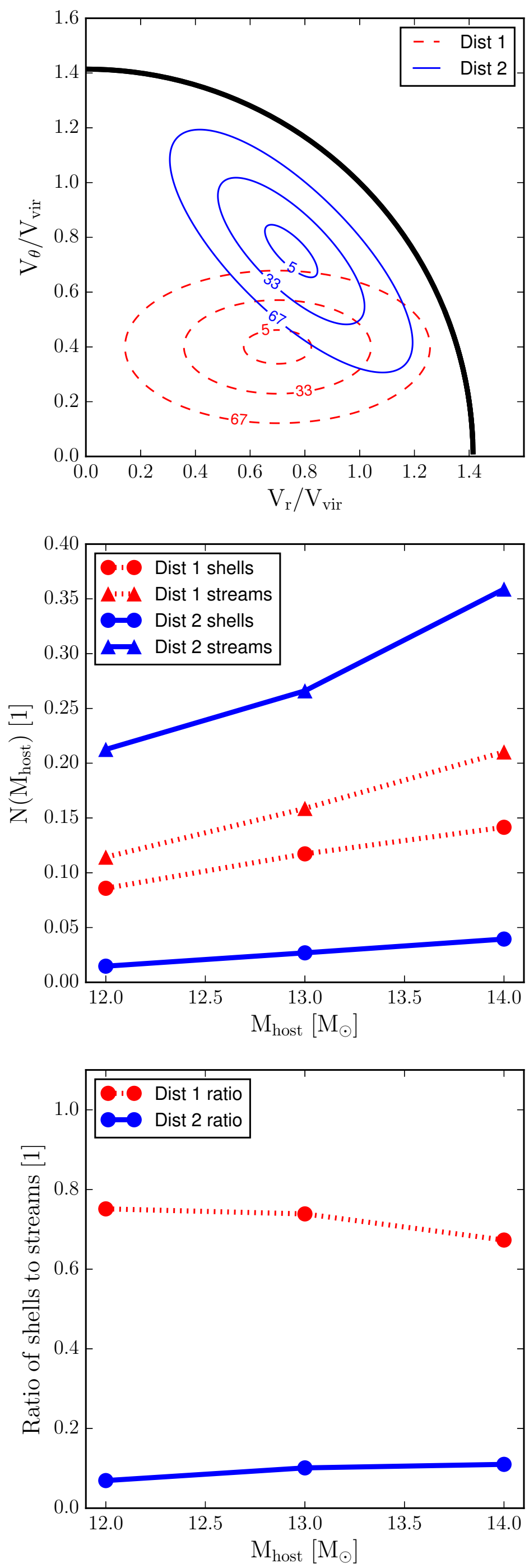

**Figure 10.** Computed ratios of shells and streams given two simple Gaussian infall distributions. Top: contours containing 5%, 33%, and 67% of orbits for the selected infall distributions. Distribution 1 is significantly more radially biased than Distribution 2 and has lower average energy. Center: Computed fraction of galaxies that would be observed to host shells or streams given the above distributions. As expected, the more radial orbits of Distribution 1 produce more shells. Bottom: the ratio of the number of shell-hosting to stream-hosting galaxies varies dramatically with the avaliable orbital parameters, from approximately 3:4 for Distribution 1 to nearly 1:10 for Distribution 2.

In this scenario the classical intuition that a more radial population of satellite orbits will produce more shells clearly holds true, as shown in the center panel of Fig. 10. The ratio of the expected number of galaxies hosting shells to those with streams for the relatively eccentric satellites of Distribution 1 is $N_{\rm shell}/N_{\rm stream} = 75\%$ for a host mass of $10^{12}{\rm M}_\odot$, decreasing to 67% for a host mass of $10^{14}{\rm M}_\odot$; shells contribute significantly to the global population of debris structures. Conversely, when the satellites are given a more equitable distribution of radial and tangential velocities as in Distribution 2, streams dominate the substructure population by nearly 10 to 1.

This experiment give credence to the idea that the debris structure population contains extractable information about the way satellites are accreted and enabling the exclusion large classes of infall distributions. In the following sections we consider distributions as measured from within a single cosmological model; while the ability to distinguish between such similar distributions is more speculative, these examples provide a test of the precision necessary from both observations and simulations to provide more detailed constraints on orbit distributions.

#### *4.3.2 Comparing infall distributions as measured in cosmological simulations*

In Fig. 11 we compare the number of debris structures of each type expected from the Benson (2005) distributions as measured at $z = 0$ and $z = 1$ (i.e. B0 and B1), which are shown in the left and center panels, respectively. Both have the form of a two-dimensional Maxwell-Boltzmann distribution in tangential velocity with a Gaussian distribution of radial velocities but the mean and dispersion in the $V_r$ Gaussian is a function of $V_\theta$ to accommodate covariance between the velocities. Our experiment asks how different the number of structures at $z = 0$ would be if the infall distribution had frozen as it was at $z = 1$ instead of evolving into B0. For B1 the covariance between $\rm V_r$ and $\rm V_\theta$ is somewhat weaker and the distribution overall is more sharply peaked at lower values for both velocities.

The right panel of Fig. 11 shows the computed values of $N_{\rm shell}(M)$ and $N_{\rm stream}(M)$ for ${\rm M_{host}/M_\odot} \in \{10^{12}, 10^{13}, 10^{14}\}$. Because B1 is more strongly peaked, a larger fraction of orbits are bound and therefore the total number of debris structures is 11% larger. While in both cases the structures are dominated by streams, the lower mean tangential velocity of B1 produces many more shells; averaging over the mass bins, there are 38% more shells but only 7% more streams in this case. It is encouraging that quite subtle differences in infall distributions actually measured in simulations result in different shell fractions relative to each other and to the number of streams that are expected to develop simultaneously.

#### *4.3.3 Mass-dependent infall*

As noted in Section 4.1.3 above, some authors have found that $\rm P(V_r, V_\theta)$ varies as a function of host halo mass. In particular, Jiang et al. (2015) fit the infall distribution in a $3 \times 3$ grid of host masses and mass ratios using a Voigt profile in total velocity and an exponential in radial velocity; we transform to the $(V_r, V_\theta)$ plane and show the results for the highest mass ratio bin, $0.05 < \xi < 0.5$, in Fig. 12.

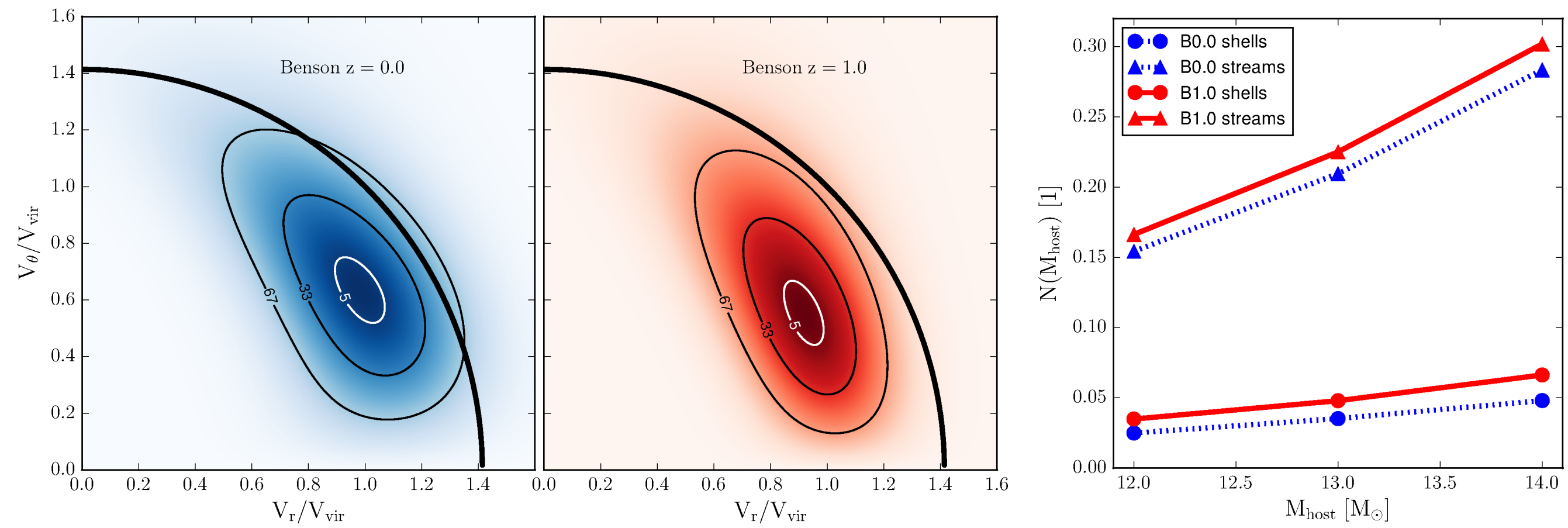

**Figure 11.** Left: visualization of the B0 infall parameter probability density. The black line is the host escape velocity while the color and contours encode the probability of an infall at a particular point in this space. Center: same but for B1. Note the lower average total velocity and the sharper peak. Right: Number of shells and streams as computed from Equation 9 using the limits described in the text. While streams (triangle markers) dominate, the number of shells (circle markers) increases 38% when changing the infall distribution from B0 (blue dashed lines) to B1 (red solid lines).

To check whether mass dependence will alter $N_{\rm shell}$ and $N_{\rm stream}$ we compute them using two scenarios: first assuming that all host haloes, regardless of mass, receive satellite infalls as described by the $10^{12}{\rm M}_\odot$, high-mass-ratio distribution (J12, leftmost) and secondly assigning them their actual host-mass-dependent, high-mass-ratio orbital infall distribution (JC, the three distributions shown).

Since the largest mass ratio mergers are more radially biased (compare also Tormen 1997; Vitvitska et al. 2002) the fractional contribution of shells to the debris population is much larger for both J12 and JC ($N_{\rm shell}/N_{\rm stream} \sim 40-60\%$) than for B0 or B1 where all values of $\xi$ are considered together, which instead results in $N_{\rm shell}/N_{\rm stream}$ of only $\sim 15-20\%$. Between J12 and JC, the $10^{12}{\rm M}_\odot$ values are identical by construction, however the differences between the $10^{12}{\rm M}_\odot$ and $10^{13}{\rm M}_\odot$ infall distributions increase the number of shells calculated around the larger host by 29 per cent while negligibly altering the number of streams.

## 5 DISCUSSION: OBSERVATIONAL PROSPECTS

The estimates of the frequency of shell-like and stream-like debris signatures in Section 4 suggest that a survey that can accurately measure the prevalence of streams and shells would be capable of providing constraints on orbital infall distributions. There is one observational sample that is interesting to compare with our preliminary estimates. Atkinson, Abraham & Ferguson (2013) surveyed 1700 galaxies to 27.7 mag/arcsec$^2$ in the $g'$ band, searching for evidence of tidal debris. Approximately 1 in 6 galaxies had strong indications of some type of tidal feature allowing percent-level constraints on the fraction of galaxies containing debris of several morphological types, as sorted by visual classification.

We can make a preliminary comparison between their results and our predictions based on the Jiang et al. (2015) infall distributions (Fig. 12). We simplify their six morphological categories by combining 'shells' and 'fans' into a single population along with 'streams' and 'linear' features, to better match our dichotomy. 'Diffuse' and 'arm' structures are presumed to be highly mixed remnants and parts of the host galaxy, respectively, and therefore we exclude them from our accounting to match the analysis above. By assuming that the observed galaxies follow the stellar-to-halo mass relation found by Mandelbaum et al. (2006) and combining the observed stellar masses into 0.4 dex bins, the Atkinson, Abraham & Ferguson (2013) results suggest that at host halo masses of $10^{12}{\rm M}_\odot$ approximately 3 per cent of the observed galaxies have shells and 6 per cent have streams, while of those nearer $10^{13}{\rm M}_\odot$ 8 per cent have shells and 10 per cent have streams. Comparing with Fig. 12, one sees that the analytic estimates overpredict both types of debris structures by a factor of $\sim 2$. In addition, the ratio of shells to streams is higher in the observations and is more strongly dependent on mass. While it may be tempting to interpret these numbers as indications that the model merger rate is too high with orbits that are excessively tangentially biased, several questions remain. These differences could instead be attributed to the simplifications made in the integration limits that were chosen to mimic a surface brightness limits in the observations or in differences between their visual classification of debris types and our analytic criteria for splitting shells and streams. More carefully matching the surface brightness cut to the survey parameters will allow us to probe these questions.

It is clear that extensive work is necessary before a conclusive comparison between observational and simulation can be made. This requires (i) an unbiased, uniform survey containing at least a few hundred galaxies with detected debris structure, such as the Atkinson, Abraham

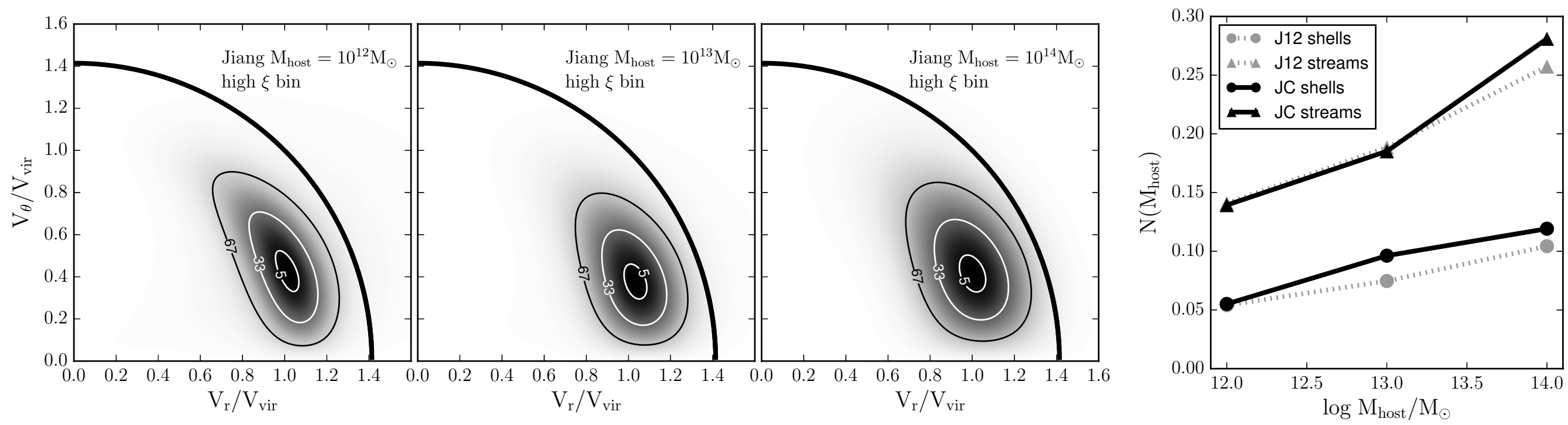

**Figure 12.** Infall distributions and resultant average number of shells and streams from J12 and JC. Left panels: The transformed velocity distributions in the highest mass ratio bin from Jiang et al. (2015). There is a trend of decreasing covariance between radial and tangential velocity with increasing host mass as well as lower average tangential velocity at larger host masses. Right: the resultant debris populations, first when assuming all hosts receive infalls according to the lowest host mass infall distribution (J12, gray lines) and then assigning each their measured probabilities (JC, black lines). The same values are found in each case at $M_{\rm host} = 10^{12} M_\odot$ by construction.

& Ferguson (2013) sample; (ii) a set of satellite disruption simulations with different mass ratios and orbits type, realistically-embedded stellar distributions, considering all viewing orientations and including contamination from the host galaxy's light; and (iii) an objective and automated method for classifying features to a given surface brightness limit in both simulations and observations in the same way.

Looking to the future, the Large Synoptic Survey Telescope will probe a region $\sim 20000$ deg$^2$ and ultimately reach a surface brightness sensitivity of $r \sim 29$ mag arcsec$^{-2}$ (Ivezic et al. 2008). Such a deep, uniform survey will unveil vast numbers of substructures. This prospect provides strong motivation for addressing the second and third requirements so that the orbits of infalling structures can be recovered.

## 6 CONCLUSION

In this work we have studied the formation of the major morphological classes of tidal debris from minor galactic mergers - shells and streams - and found that a simple physical model of stream production based on offsets in energy between the unbound particles and the progenitor satellite is extensible to shells if the consequences of angular momentum offsets are also included. This is quantified through a morphology metric which is shown to provide a good match to visual classification. For an individual merger the resulting morphological class is intrinsically time-dependent because the energy effects are bounded by the size of a single orbit but those of the angular momentum are not. We demonstrate how the distribution of merging satellite orbital parameters measured in cosmological N-body simulations can be interpreted through a morphology metric as predictions for the fraction of galaxies that have experienced stream- or shell-creating mergers. Our results show that different infall distributions produce results that are plausibly distinguishable using studies of low-surface-brightness features around nearby galaxies. Tidal debris morphology thus provides unique access to orbit distributions - an as yet unexplored part of the accretion history.

## ACKNOWLEDGMENTS

We thank Jacqueline van Gorkom, Robyn Sanderson, and Adrian Price-Whelan for useful comments and discussions and the anonymous referee for a constructive review. The authors acknowledge the Shared Research Computing Facility (Yeti[2]) at Columbia University for providing HPC resources that have contributed to the results reported in this paper. This work was supported by NSF grant AST-131296.

[2] http://hpc.cc.columbia.edu/